# A Global Citizen of the Skies

by

Dr. Daniel Brown

*School of Science and Technology*

*Nottingham Trent University*

*Clifton campus*

*Nottingham*

*NG11 8NS*

*Telephone: 0115 848 3518*

*email: daniel.brown02@ntu.ac.uk*

&

Natasha Neale

*The Centre for Effective Learning in Science*

*School of Science and Technology*

*Nottingham Trent University*

*Clifton Lane*

*Clifton*

*Nottingham*

*NG11 8NS*

*Telephone: 0115 848 8065*

*email:natasha.neale@ntu.ac.uk*

**Abstract**

Global citizenship plays an important role in today's schools. Many subjects taught in schools have already incorporated such ideas. Science and physics have also followed suit. However, when dealing with astronomy - a topic so seemingly far removed from society - it becomes difficult to imagine any links with global citizenship.

At Nottingham Trent University observatory we have developed an activity ideal to transport the idea of global citizenship and deal with common astronomical misconceptions. It incorporates role playing in the inspiring environment of an astronomical observatory.

In this activity stellar constellations with their myths and history pose an ideal opportunity to explore global citizenship. Students not only place themselves in someone else's situation, but also compare their different reactions when faced with a common situation. This idea is extended to analyse the change in constellation culture throughout time and the affect politics has had on constellations.

In this article we outline the details of this activity and how it deals with common astronomical misconceptions. We also demonstrate its implementation into astronomy taught at schools. First results showing the impact the activity had on the students will also be given.

**Introduction**

In the past decade citizenship has increasingly become more important for the schools in the UK and has lead to it being integrated into the national curriculum. Citizenship covers key concepts such as: democracy and justice, rights and responsibilities, as well as identities and diversities. Furthermore, the Department for international Development has also given schools a guideline to explore their 'global dimension' (Department for International Development, 2005). This new dimension is compatible with the goals of citizenship since it contains the important ideas of values and perceptions. Moreover, we are all citizens of planet Earth. Therefore, the global dimension of citizenship also needs to be included into the teaching.

However, Andrews & Mycock (2007) describe how schools struggle to provide adequate citizenship lessons. They state that there still remains confusion in how to deliver the key concepts taught in this topic. Although we cannot resolve these problems we can offer an exciting and versatile approach to global citizenship from the perspective of astronomy in the science curriculum.

Astronomy is a specific branch of science dealing with planets, stars, our Galaxy, and the Universe in general. It spans distances of billions of kilometres, too gigantic to even imagine. It covers times of billions of years, well beyond the normal comprehension. Therefore, it seems to be difficult to link this subject with any issues arising from global citizenship and to everyday life. But this branch of sciences and its amazing images has always inspired children and made them curious of the world that is out there. Here already lies the ideal foundation to link astronomy with global citizenship. Even in a more formal structure of the key concepts given in the national curriculum, the concept of 'Cultural understanding' in science - covering how science and here astronomy has its roots in many different societies and cultures - slots in nicely into the previously mentioned concept of values and perception in citizenship.

Utilising astronomy to convey global citizenship follows the advice formulated by Ofsted (2003) by not only providing community service opportunities but applying citizenship in a more imaginative and practical way. Furthermore, the presented activities encourage tolerance and mutual understanding that Miller (2000) claims define citizenship education.

**Where does astronomy 'fit' into the national curriculum?**

The national curriculum has a section devoted to astronomy subjects throughout key stage 2 containing a section covering 'The Earth and beyond'. Key stage 3 has the 'Environment, Earth and the Universe' and key stage 4 'Environment, Earth and Universe'. The exception is key stage 1 where there is no devoted section to objects in space and the observation there of (National Curriculum 2009). However, it could be argued that the concept of forces used at key stage 1 can be applied to the motion of objects in orbit around our Sun when describing gravity.

The subject is defined at key stage 2 under 'The Earth and beyond' title by teaching the students that;

- The Sun, Earth and Moon are approximately spherical.

- How the position of the Sun appears to change during the day, and how shadows change when this happens.

- How the day and night are related to the Earth's spin on its own axis.

- That the Earth orbits the Sun once a year and that the Moon takes approximately 28 days to orbit the Earth.

Key stage 3 pupils cover the above mentioned points in more detail and build on the knowledge gained at key stage 2. Under this section students are taught '... astronomy and space science provide insight into the natural and observed motions of the Sun, Moon, stars, planets and other celestial bodies ...' (National Curriculum 2009b). The progression to key stage 4 sees the first introduction to the solar system and pupils will discover that the solar system is part of the Universe, which has changed since its origin and continues to show long term changes (National Curriculum 2009c).

Previously, the inclusion of astronomy has been firmly fixed in the national curriculum for science and no other subject. Since September 2002 however, the science curriculum has become less focused on just learning the facts and more open to discovering how the science can be applied by the introduction of a mandatory programme for the study of citizenship (Warren et al. 2005). Throughout all the key stages the core science now runs parallel to a heavy emphasis on investigating 'how science works' and the relevance of it to everyday life.

The citizenship curriculum allows the growth of transferable skills and the essential science skills such as effective communication by both traditional and new media, investigations and experimentation as well as research methods. This new emphasis on 'how science works' is applied throughout all key stages but in more detail at the higher key stages. Through this part of the curriculum, students are expected to be able to appreciate and discuss the tentative nature of scientific understanding and knowledge. They should be able to discuss and argue the risks, benefits, cultural, social and ethical issues surrounding science as well as appreciate how scientific knowledge is used by society to inform decision making.

There has been an increasing movement by examining bodies to broaden the qualifications available to students with particular interests. This has led to the introduction of GCSE qualifications not just in Physics with a module on astronomy but a whole GCSE in the subject (National Database of Accredited Qualifications 2009). A-level qualifications are also available on astronomy as a standalone subject. The introduction of these science disciplines to the national curriculum has seen an increasing number of teachers needing to update the knowledge they have of astronomy.

**Who is Teaching Astronomy in Schools?**

It has long been recognised that primary level teachers (that is to say those dealing with the teaching of key stages 1 and 2) have always needed to have a broad subject knowledge since they are responsible for teaching the class and not just a subject within the school. Primary schools tend to have subject coordinators on site, usually a teacher with that subject background, to assist in the delivery of topics that fellow teachers with a different background may not feel overly confident in delivering. Secondary schools do not have this luxury and teachers need to be subject specific and are expected to be able to cover all aspects of that subject as required under the national curriculum guidelines. National shortages in the numbers of secondary school science teachers has lead to the introduction of government schemes backed by teacher training bodies to encourage more science graduates in particular into secondary education. These are usually monetary incentives to teach such as the tax free bursary and golden hello schemes (TDA 2009). There is a national trend that more science graduates have a qualification in a biological science than either chemistry or physics. This impacts on the teacher training courses as this unbalance of numbers also leads to an unbalance of trainee / newly qualified teachers in either chemistry or physics.

There have been numerous studies undertaken that highlight this problem well. It has been shown that the number of secondary school physics teachers that do not have a qualification at degree level in physics is 66 %. Furthermore, the number of teachers teaching physics who do not have an A-level qualification in the subject is 29 % while these numbers are reduced for the biosciences in particular. 39 % of those teaching key stage 4 students do not have a degree level qualification in a bioscience subject (Council for Science and Technology 2000).

The Centre for Effective learning in Science at Nottingham Trent University runs a wide range of STEM (science, technology, engineering and maths)outreach activities with pupils from all backgrounds and across all key stages. Discussions with teachers across all key stages has emphasised that they are more comfortable in involving their students in hands-on practical science when dealing with the subject they have a background in. Evaluations of activities delivered by the Centre for Effective Learning in Sciences (CELS) with pupils across all key stages has highlighted that the most enjoyable aspect of the science they do in schools are the practical activities.

It is for these reasons that there is a great potential for outreach activities and for teacher continuing personal development sessions to enrich the science experience of the pupils by studying astronomy. Teachers from all science backgrounds can be encouraged to consider how astronomy, the discoveries already made as well as those yet to be made fit into the citizenship curriculum.

Citizenship allows pupils of all abilities to engage in the topics covered. Lower ability pupils as well as higher ability pupils can tackle topics together in a less academically structured way.

The outreach programme developed at the Centre for Effective Learning in Science has taken on board recommendations laid down by HEFCE (Higher Education Funding Council for England, 2007) . The Centres policy of making science relevant, accessible and achievable has ensured that we have taken into account the need to supply information in a range of formats for it to be most readily accepted by the majority of students. We have also allowed room for the students on the activity to interact with each other and learn from each other.

**The Stellar Constellation Activity**

The stellar constellation activity lasts 45 minutes and involves two groups of 6 pupils role playing two groups of people from a different cultural background (in the following called the 'Explorers' and the 'Natives'). To prime both groups, they are provided with a background story on what they are representing supported by a role play involving the use of artefacts such as a sextant, candles, and flashlights. Both groups meet in the environment of the Nottingham Trent observatory planetarium that is showing a night sky with stars. They then separately develop stellar constellations and stories about them. To aid their work, we provide them with a simplified star chart of the stars they can see to draw their constellations and write stories about them. These handouts are themed towards either the 'Explorers' or the 'Natives' reminding them of their role. The results are then presented to each other and discussed, with respect to how useful the other cultural group would find the new constellations. A supervisor for each group is always present to facilitate the activities and stimulate discussion.

The activity then ends with a brief presentation on the Greek creation story of the constellation of Orion. Furthermore, the asterism of the Plough is introduced and how e.g. the slaves escaping to North America used it to guide their way. The entire activity has been designed to cater towards a range of different learning styles applied by the participants, especially role play which is still rarely included into science teaching. This is an obvious missed opportunity as work by Francis (2006) has shown how powerful role play can be in developing problem solving and empathy skills.

The impact of the stellar constellation activity is determined using a pre- and post-assessment questionnaire that has been modelled to some extent on the 'Astronomy Diagnostic Tool' developed by the 'Collaboration for Astronomy Education Research' (Brogt et al. 2007). Given the age group of the participants, we support their work with the questionnaire in such a way as not to influence their decisions. We have developed the questionnaire to cover areas of common astronomical misconceptions on the topic of constellations as well as focusing on the aspects of citizenship. The entire questionnaire is given in the appendix. The following 4 questions cover astronomical misconceptions that we have encountered during our previous work with secondary and undergraduate students:

- Q1: *Which of the following do you think could be a stellar constellation?*

    If constellations are known, they are only supposed to have been invented by a western/middle eastern culture. That other cultures had their own take on stellar constellations is often overlooked.

- Q2: *What is so special about the Pole Star?*

    It is still believed that the Pole Star is the brightest star in the Sky. However, it is only the 49$^{th}$ brightest star.

- Q3: *Which of the following stars belongs to the constellation of Orion?*

    Constellations are thought to be only groupings of stars. They are very much more like official borders in the sky.

- Q4: *How far away do you have to travel to see the change in the shape of the Plough shown here?*

    Stellar constellations are often thought to be two dimensional. However, they are only the two dimensional projection of a three dimensional distribution of stars.

Global citizenship issues are targeted in the following longer three questions:

- Q5: *You tell your German pen friend – who is also interested in astronomy – that you have seen the Plough in the Sky last night. He has no clue what you are talking about. Could this be because: ...*

    In this question we explore the cultural background of other countries compared to our own.

- Q6: *If you were a Viking living in Britain 1000 years ago, how would the stellar constellations you know be compared to today's constellations?*

    Here we develop an understanding that the UK has never been an isolated community. It demonstrates how even staying in your own country and cultural region, your culture has evolved over time.

- Q7: *You are a young astronomer 300 years ago. Your first job for your King is to draw a map of the stars including constellations. How would you make your King happy and keep your new job?*

    Apart from cultural influences, political pressure can also lead to changes in the perception of e.g. stellar constellations. Here the wider analysis of a situation is promoted, taking into account political constraints.

**Results**

The activity has only been running for 6 months including just over 100 participants ranging in ages from 9 years up to PGCE primary science students. During the activity we have been not only collecting the responses given in the questionnaires, but have also saved a sample of constellations created by some of the participants as well as feedback provided after the

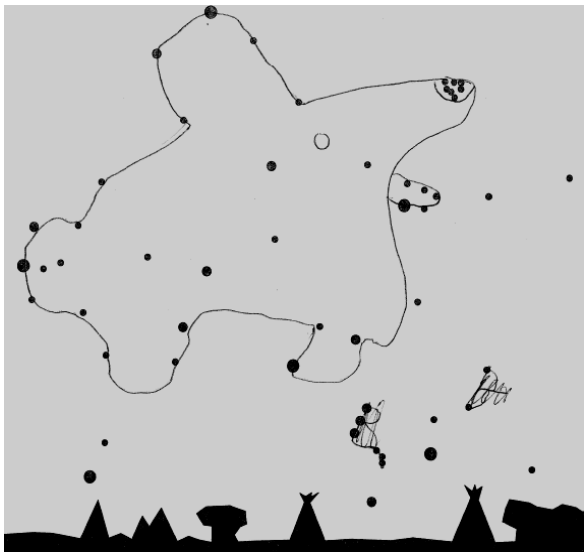

**Figure 1** – A stellar constellation called 'The Sun Rabbit' and created by a participant in the 'Natives' group.

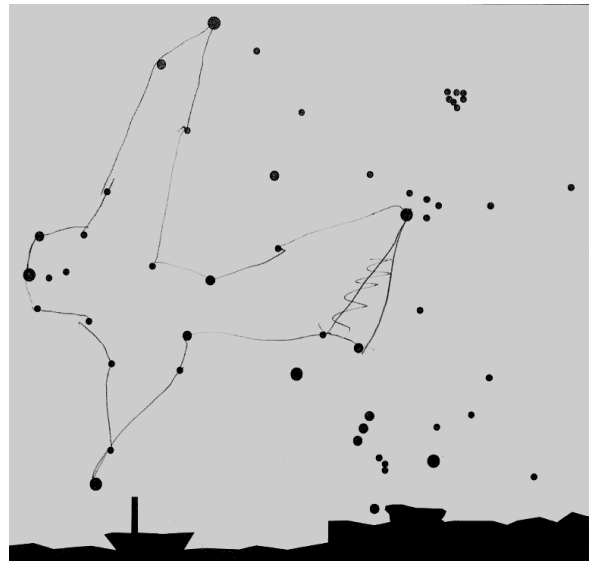

**Figure 2** – A stellar constellation called 'The Plane' and created by a participant in the 'Explorers' group.

Figure 1 and 2 show two examples of stellar constellations created during the activity a member of the 'Natives' and 'Explorers' group, respectively. Already a clear impact of the cultural background of each group can be detected. The 'Natives' constellation targets an animal rather than an artificial object like a plane chosen by the member of the 'Explorers'. Even though the 'Explorers' group were intended to be based in the times of Columbus and a plane is clearly an unknown device in these times, placing such technology into the sky demonstrates the importance it has for this group. The 'Natives' on the other hand, are far more orientated towards nature that is especially highlighted in the story written about the 'Natives' constellation:

"*The rabit in the sky waches over the natives of America, keeps the fires burning and the water running rolls the sun out of the borrow every day and back at night and vice-versa with the moon.*"

These results are representative of the success the role playing activity had in placing the members of the group into the right set of mind. During the discussion of the constellations of each group the participants clearly saw the value of their constellations compared to the constellation of the other group. Both groups were surprised about how different their constellations were, seeing that they were based on exactly the same stars. After the activity one participant mentioned: "*I never knew stars could be that interesting*".

Apart from these subjective results, we have also analysed and compared the results of the pre- and post-assessment questionnaires. We will only present a selected group of questions shown in Fig. 3-6. All figures show the frequencies in percent of the answers numbered one to five (see appendix for answers). The pre-assessment results were shaded in gray compared to the post-assessment results that are shaded in black.

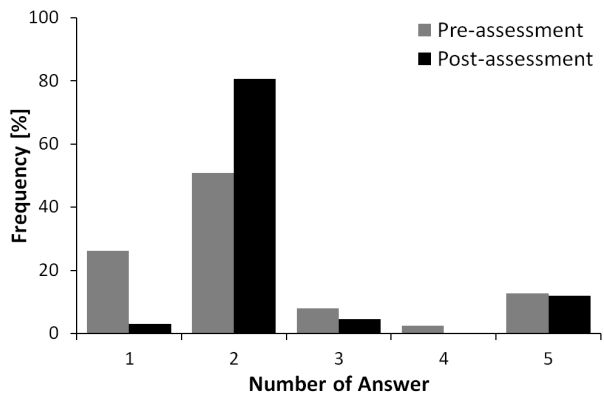

**Figure 3** – The distribution of the answers in percent to the Pole Star question (Q2). Pre- and post-assessment results are displayed as given in the legend. Prior to the activity, a significant number of participants thought the pole star was the brightest star (answer 1). After the activity this fraction was clearly reduced and the finding north option (answer 2) was favoured.

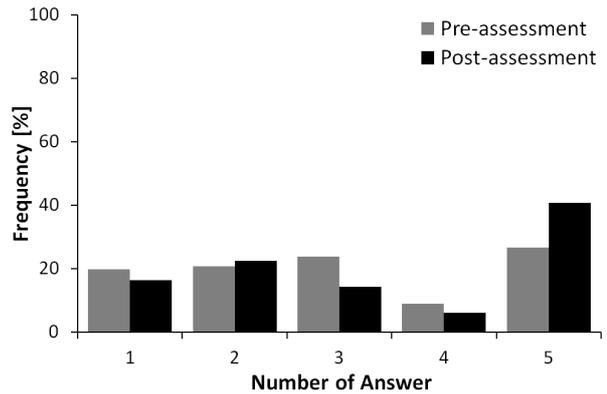

**Figure 4** – The distribution of the answers in percent to the change in a constellation question (Q4). Pre- and post-assessment results are as in Fig. 3. No clear signal prior to the activity can be detected. After the activity both the distant star (answer 2) and the America (answer 5) answer option increase, with the America option as the most dominant response.

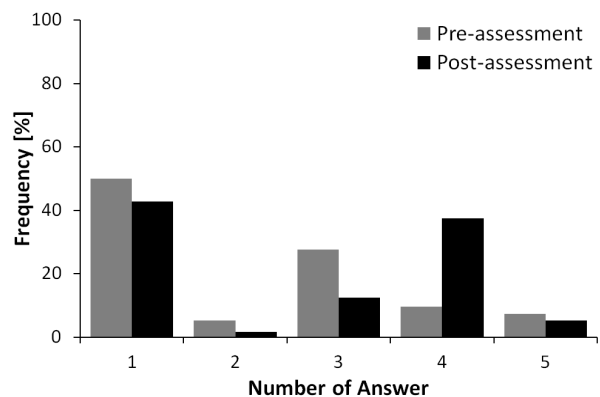

**Figure 5** - The distribution of the answers in percent to the stories about constellation question (Q6). Pre- and post-assessment results are as in Fig. 3. The responses prior to the activity all use the apparent location of the stars to justify their answer (answer 1 & 3). However, after the activity the answer relating to our change in culture (answer 4) has clearly increased.

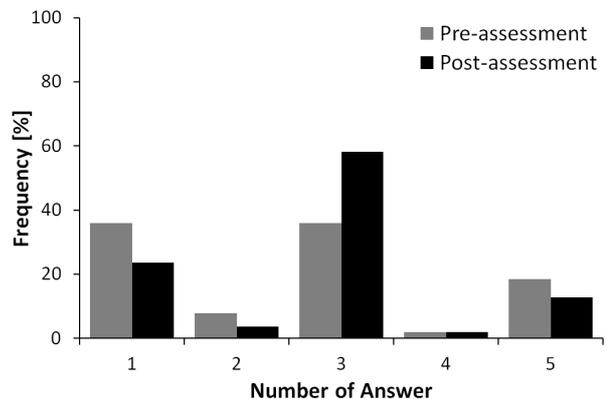

**Figure 6** – The distribution of the answers in percent to the creating constellation map question (Q7). Pre- and post-assessment results are as in Fig. 3. Pre-assessment shows how important the precision (answer 1) and copying already existing maps (answer 5) are. However both responses are reduced after the activity, in favour of creating a constellation more appropriate to the political conditions (answer 3).

We will briefly discuss each of these questions.

The results on the question targeting the pole star (Q2) are presented in Fig. 3. Although a significant number of participants already knew about the use of the pole star as a pointer towards North (answer 2), more than a quarter thought it was also the brightest star in the sky (answer 1). After the activity this fraction decreased dramatically to less than 3% and nearly all of the participants understood the use of the pole star to find north.

Fig. 4 shows the results on the change in a constellation question (Q4). No clear preferred response can be detected in the pre-assessment. After the activity the responses have changed marginally, with the preferred answers being America (answer 5) followed by 'A distant Star' (answer 2).

The findings on the question targeting the stories we tell about stars (Q6) are presented in Fig. 5. Both preferred answers use the actual location of the stars as justification, either the stories stay the same (answer 1 - 50%) or change (answer 3 - 28%) due to the same or different patterns of the stars. After the activity the frequency of only the response using the change in the stories we tell about the stars as a justification (answer 4) increases to 38%. The most frequent answer still remaining with 43% is that the stories remain the same, since the stars do not move.

The results on creating a map of the stars question (Q7) are presented in Fig. 6. In the pre-assessment there are three preferred answers. They include taking care of high precision (answer 1) and repeatability (answer 5) but also including a constellation named after the King (answer 3). After the activity the latter response has increased to the most popular answer with a frequency of 58%.

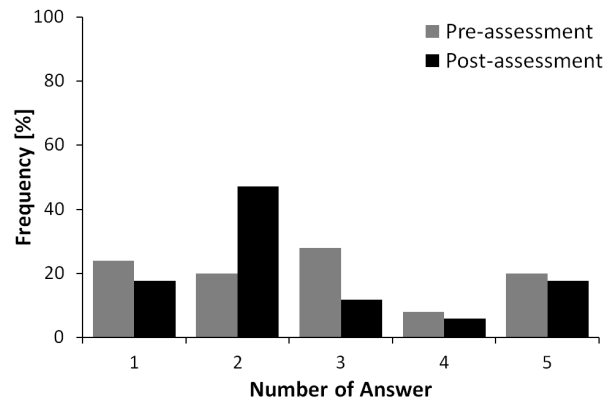

**Figure 7** – The distribution of the answers in percent to the change in a constellation question (Q4) for only Primary science PGCE students. Pre- and post-assessment results are as in Fig. 3. Even in this sub sample, no clear prior knowledge of the 3 dimensional structures of stars in our galaxy can be detected. After the activity, their response clearly changed in favour of the distant star answer (answer 2).

Taking into account all these results we can demonstrate how successful our stellar constellation activity was in terms of correcting common astronomical misconceptions, e.g. concerning the pole star. Only a very small fraction of participants thought of it as the brightest star afterwards. However, the activity failed to demonstrate how stellar constellations are a result of the three dimensional structure or the stars and the length scales involved. We seemed to have even created a misunderstanding that America, the home of our 'Natives' group, is the typical distance over which to see a geometric change in the shape of the constellations. We selected a sub-sample only consisting of 60 PGCE science primary students (results are presented in Fig. 7 in a similar manner than in Fig. 3) to analyse if the responses change when targeting an older age and more scientific orientated audience. To our surprise we found again no clear response in the pre-assessment concerning the change of a constellation due to the position of the observer. However, they did change their responses after the activity towards the correct answer.

The activity has been a full success on transfering relevant skills related to global citizenship. The participants have fully understood how it is the culture itself that changes constellations, shown in the responses to Q6 and Q7. This is comparable to understanding how our culture influences many other aspect of society. It even allows the participants to experience for themselves how under the same conditions another culture reacts differently. Furthermore, the participants have also gained an understanding of the political influence science and a community in general experiences. In Q7 more than half of the participants would create a constellation purely to please their King, rather than follow scientific precision and the general consensus.

**Conclusions**

Given our preliminary findings we intend to improve our activity in three ways.

Firstly, we intend to increase the significance of our findings by extending our sample beyond the 100 participants. This will ensure that some of our findings that show an unclear response pattern (e.g. Q1, Q3 and Q5) become clearer and allow us to analyse the data in more depth by selecting a sub-sample as in the case of Q4.

Secondly, we intend to optimise the questionnaire using our preliminary findings. Such work is already in progress. We have already modelled the creation of a stellar constellation on the handout, increased the size of the handout from A4 to A3, and using pencils rather than pens. These steps have already helped to engage the participants more to the activity since it stimulates group work and eliminates the fear of committing themselves to a certain constellation. (Note the scribbles in Fig. 1 and 2 trying to eliminate some unwanted constellations.) We further intend to establish an interview with a selected focus group to analyse the reasoning behind their answers. This interview will allow us to test the effectiveness of the questionnaire and eliminate unintended responses. Such a glitch is the incorrect association of America as a preferred location to observe a change in the shape of a constellation. This was influenced by the choice of the cultural background of the 'Natives'.

Thirdly, we would like to follow up certain groups of participants, e.g. PGCE primary science students and inquire what impact if any the activity has had in their teaching of science and incorporating global citizenship.

We conclude that astronomy only seems to be a science far removed from society. The skies above us unite many different cultural regions and countries across the world and enabling us to explore the global citizenship dimension within astronomy. This activity is only one possibility in incorporating the key concepts of citizenship within a lesson. It also allows the students to apply their different learning styles and discover these key concepts by themselves.

As further consequence of our findings, we have unearthed the need for more emphasis on physics and especially astronomy when training new science teachers.

The basic idea of length scales when dealing with stars and stellar constellations is not only a specific skill in astronomy and should be fully embraced by science teachers. We are providing resources to run the stellar constellation and material to extend this activity on the following CELS website:

**http://www.ntu.ac.uk/cels/outreach/Optical_observatory/Projects/index.html**

# Stellar Constellation Questionnaire

You can answer each question with one or more choices.
Mark your choice of **answer** or **answers** for each question!

Q1: Which of the following do you think could be a stellar constellation?

1. Sword
2. Leo
3. Blubber Container
4. Plough
5. Laboratory Furnace

Example of a stellar constellation

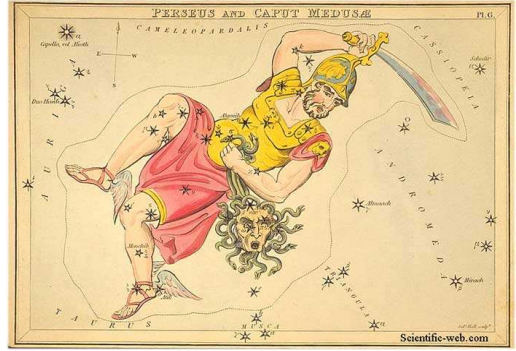

Q2: What is so special about the Pole Star?

1. It is the brightest star
2. It helps you to find north
3. You can only see it at the Poles
4. It appears when winter comes
5. It never moves

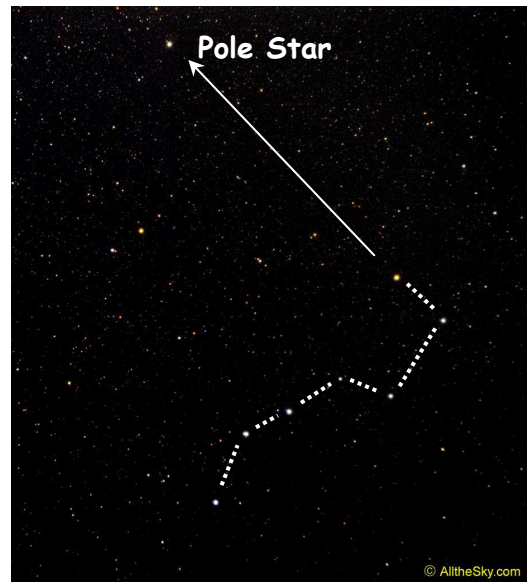

Q3: Which of the following stars belongs to the constellation of Orion?

1. Star A
2. Star B
3. Star C

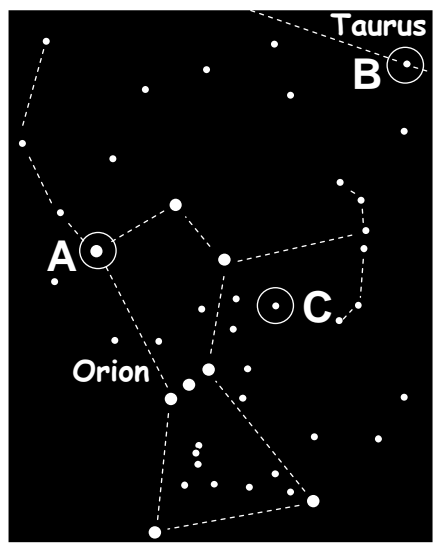

**Q4: How far away do you have to travel to see the change in the shape of the Plough shown here?**

1. Across the country
2. A distant star
3. Moon
4. Pluto
5. America

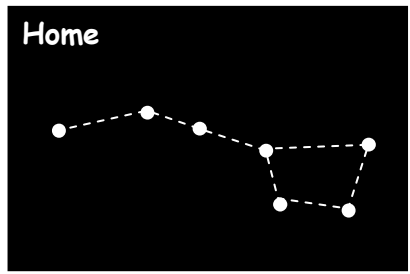
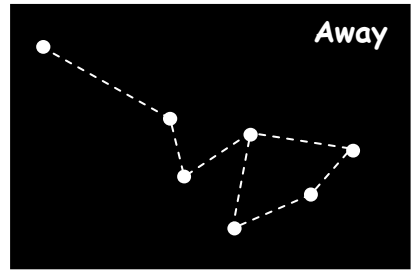

**Q5: You tell your German pen friend - who is also interested in astronomy – that you have seen the Plough in the Sky last night. He has no clue what you are talking about. Could this be because:**

1. He can't see the Plough in Germany when you can in England
2. He can't see the Plough in Germany at all
3. He calls the same stars the ´Large Carriage´
4. Each star in the Plough is at a different position in Germany
5. In Germany nobody gives these stars names

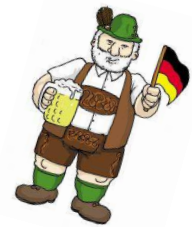

**Q6: If you were a Viking living in Britain 1000 years ago, how would the stellar constellations you would know be compared to today's constellations?**

1. The same, because the stars don't move
2. The same, because the stories you tell about the stars don't change
3. Different, because each star is at a different location in the Sky
4. Different, because the stories you tell about the stars are different
5. Different, because you can't see any of the stars you see today

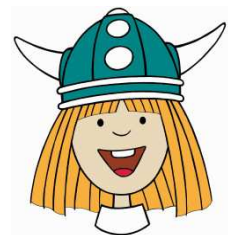

**Q7: You are a young astronomer 300 years ago. Your first job for your King is to draw a map of the stars including stellar constellations.
How would you make your King happy and keep your new job?**

1. Take care to exactly measure the stars' position
2. Write something in the map thanking him for choosing you for the job
3. Create a new stellar constellation *The best King ever* and name it after him
4. Use the most expensive materials
5. Use the same constellations than everyone before

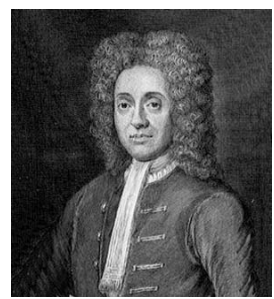